\documentclass{article}

\usepackage[hidelinks]{hyperref}
\usepackage{amsmath,amsfonts,bm,mathrsfs,amssymb,amsthm}
\usepackage{nicefrac}
\usepackage{microtype}
\usepackage[dvipsnames]{xcolor}
\usepackage{graphicx,subcaption}
\usepackage[inline]{enumitem}
\usepackage[top=1.25in, bottom=1.25in, left=1.25in, right=1.25in]{geometry}

\theoremstyle{remark}

\newtheorem{example}{Example}

\DeclareMathOperator{\tr}{tr}
\def\given{\,|\,}

\title{Robust Bayesian inference in complex models with possibility theory}

\author{%
  Jeremie Houssineau \\
  Department of Statistics\\
  University of Warwick\\
  \vspace{.5em}
  \texttt{jeremie.houssineau@warwick.ac.uk} \\
  \vspace{.5em}
  and \\
  David J.~Nott \\
  Department of Statistics and Applied Probability\\
  National University of Singapore \\
  \texttt{standj@nus.edu.sg} 
}

\date{}

\begin{document}

\maketitle

\begin{abstract}
We propose a general solution to the problem of robust Bayesian inference in complex settings where outliers may be present. In practice, the automation of robust Bayesian analyses is important in the many applications involving large and complex datasets. The proposed solution relies on a reformulation of Bayesian inference based on possibility theory, and leverages the observation that, in this context, the marginal likelihood of the data assesses the consistency between prior and likelihood rather than model fitness. Our approach does not require additional parameters in its simplest form and has a limited impact on the computational complexity when compared to non-robust solutions. The generality of our solution is demonstrated via applications on simulated and real data including matrix estimation and change-point detection.
\end{abstract}

\section{Introduction}

Robustness is a crucial element for bridging the gap between simulation and real data:
\begin{enumerate*}[label=\roman*)]
\item as summarised by George Box's famous aphorism ``all models are wrong, but some are useful'' \cite{box1976science}, there will always be deviations between the model and the true data-generating mechanisms, and
\item most real data sets are contaminated by outliers which become increasingly difficult to remove manually as the dimension of the data or the size of the data set becomes larger.
\end{enumerate*}
The need for robustness has long been acknowledged in the literature, starting with M-estimator \cite{huber2004robust}, and continues to be the central motivation of a large body of work, e.g.\ \cite{minsker2017robust,knoblauch2018doubly,miller2019robust,fearnhead2019changepoint,cheriefabdellatif2020mmd,boustati2020generalised,ma2020robust}. Particularly relevant are the methods of \cite{knoblauch2018doubly} and \cite{miller2019robust}, which rely on different forms of likelihood discounting, either via the use of a suitable loss function \cite{knoblauch2018doubly} or as a consequence of ``coarsening'' the conditioning in Bayes' theorem \cite{miller2019robust}. Related to robust inference is the problem of identifying inconsistencies between prior and data \cite{bayarri2007bayesian,nott2020checking,nott2021using} where it is generally the prior that is deemed unsuitable in the presence of a conflict. The objective of this work is to study the opportunities to address robust inference which are arising when viewing Bayesian inference through the lens of possibility theory \cite{dubois2015possibility}.

Bayesian inference is well known for the performance it naturally yields in terms of model selection, via the notion of evidence or marginal likelihood. The evidence assesses the tightness of a model, which combines a measure of the consistency between the model and the data with a measure of the simplicity of the model. When following the Bayesian approach in the context of possibility theory, the analogue of the notion of evidence only assesses the consistency between the model and the data; although this aspect implies that model selection cannot be carried out in the same way as in the standard Bayesian framework, it creates an opportunity to discount the data based on this notion of consistency and devise robust inference algorithms. The main advantages of this approach are that it is free from tuning parameter and that it can be applied in closed-form under conjugacy.

The relevant background in possibility theory will be covered in Section~\ref{sec:possibility_theory} before introducing the proposed approach to robust inference in Section~\ref{sec:robust_inference}. Limitations of our approach are then discussed in Section~\ref{sec:limitations} and simulations on simulated and real data are presented in Section~\ref{sec:simulations}.

\section{Possibility theory}
\label{sec:possibility_theory}

\subsection{General setting}

Possibility theory \cite{dubois2015possibility} can be interpreted as modelling deterministic uncertainty \cite{walley1991statistical} as opposed to uncertainty stemming from random phenomena. As in probability theory, we consider a sample space $\Omega$ which contains all the possible states of nature and define a (deterministic) \emph{uncertain variable} as a mapping $\bm{\theta}$ from $\Omega$ to a parameter set $\Theta$; yet, instead of endowing $\Omega$ with a probabilistic structure, we simply define $\omega^* \in \Omega$ as the true state of nature from which it follows that $\bm{\theta}(\omega^*)$ is the true value of the parameter of interest. The available information about $\bm{\theta}$ can be described by a \emph{possibility function} $f_{\bm{\theta}}$ on $\Theta$, i.e.\ $f_{\bm{\theta}}$ is non-negative and verifies $\sup_{\theta \in \Theta} f_{\bm{\theta}}(\theta) = 1$. The law of large numbers and central limit theorem for uncertain variables \cite{houssineau2019elements} motivate the following notions of expected value and variance 
\begin{equation*}
\mathbb{E}^*(\bm{\theta}) = \arg\sup_{\theta \in \Theta} f_{\bm{\theta}}(\theta) \qquad \text{and} \qquad \mathbb{V}^*(\bm{\theta}) = \mathbb{E}^*\bigg( -\dfrac{\mathrm{d}^2}{\mathrm{d} \theta^2} \log f_{\bm{\theta}}(\theta) \bigg)^{-1},
\end{equation*}
with the variance being defined when $\mathbb{E}^*(\bm{\theta})$ is a singleton and when $f_{\bm{\theta}}$ is twice differentiable at $\mathbb{E}^*(\bm{\theta})$. The expected value verifies that $\mathbb{E}^*(T(\bm{\theta})) = T(\mathbb{E}^*(\bm{\theta}))$ for any mapping $T$. These notions are local, which corresponds to the fact that we are only interested in a single point $\bm{\theta}(\omega^*)$ rather than in a whole distribution as in probability theory. In particular, $\mathbb{E}^*(\bm{\theta})$ and $\mathbb{V}^*(\bm{\theta})$ match with the mean and variance in the Laplace approximation. One important difference between possibility functions and probability density functions (p.d.f.s) is that the former are not densities; it follows that the change of variable formula for uncertain variables does not contain a Jacobian term: if $\bm{\theta}$ is an uncertain variable in $\Theta$ described by the possibility function $f_{\bm{\theta}}$ and if $\bm{\psi} = T(\bm{\theta})$ for some mapping $T$ on $\Theta$ then
\begin{equation*}
f_{\bm{\psi}}(\psi) = \sup \{ f_{\bm{\theta}}(\theta) : \theta \in \Theta, \psi = T(\theta) \},
\end{equation*}
with the assumption that $\sup \emptyset = 0$.

The \emph{credibility} of the event $\bm{\theta} \in A$ is defined via the set function $\bar{P}_{\bm{\theta}}(A) = \sup_{\theta \in A} f_{\bm{\theta}}(\theta)$ for any $A \subseteq \Theta$; it follows that $\bar{P}_{\bm{\theta}}$ is an outer measure additionally verifying $\bar{P}_{\bm{\theta}}(\Theta) = 1$, so that we refer to it as an \emph{outer probability measure} (o.p.m.). The scalar $\bar{P}_{\bm{\theta}}(A) \in [0,1]$ can be interpreted as the maximum \emph{subjective} probability that we are ready to assign to the event $\bm{\theta} \in A$. If, for a fixed $\theta \in \Theta$, $Y$ is a random variable in $\mathsf{Y}$ distributed according to a p.d.f.\ $p_Y(\cdot \given \theta)$, then, omitting measure-theoretic details, a more general o.p.m.\ \cite{houssineau2018parameter} can be defined as
\begin{equation*}
\bar{P}_{\bm{\theta},Y}(A \times B) = \sup_{\theta \in A} f_{\bm{\theta}}(\theta) \int \bm{1}_B(y)  p_Y(y \given \theta) \mathrm{d} y, \qquad A \times B \subseteq \Theta \times \mathsf{Y},
\end{equation*}
where $\bm{1}_B$ is the indicator function of the set $B$. Defining conditioning with $\bar{P}_{\bm{\theta},Y}$ in the same way as in probability theory yields the posterior possibility function \cite{chen1995statistical,walley1999upper}
\begin{equation}
\label{eq:Bayes_theorem}
f_{\bm{\theta}|Y}(\theta \given y) \doteq f_{\bm{\theta}}(\theta \given Y = y) = \dfrac{p_Y(y \given \theta)f_{\bm{\theta}}(\theta)}{\sup_{\theta' \in \Theta} p_Y(y \given \theta') f_{\bm{\theta}}(\theta')}, \qquad \theta \in \Theta,
\end{equation}
where $y$ is a given observation such that the evidence $e(y) \doteq \sup_{\theta \in \Theta} p_Y(y \given \theta) f_{\bm{\theta}}(\theta)$ is positive, and where ``$\doteq$'' emphasises that the l.h.s.\ is a notation for the r.h.s. The main differences between \eqref{eq:Bayes_theorem} and the standard Bayes theorem are that the prior is a possibility function and the denominator is based on maximisation rather than integration. These differences are small enough to allow for most of the intuition about Bayesian inference to translate to this approach and substantial enough to create opportunities at the methodological and practical levels. In particular, it is always possible to model the total absence of information a priori by considering $f_{\bm{\theta}} = \bm{1}$, with $\bm{1}$ the function equal to $1$ everywhere; as opposed to the probabilistic case, this prior is proper as a possibility function and therefore avoids the pitfalls of improper prior p.d.f.s \cite{dawid1973marginalization}. The posterior expected value defined as
\begin{equation*}
\mathbb{E}^*(\bm{\theta} \given Y = y) = \arg\sup_{\theta \in \Theta} f_{\bm{\theta}|Y}(\theta \given y)
\end{equation*}
is the maximum a posterior (MAP), which transforms coherently under re-parametrisation: if $\bm{\psi} = T(\bm{\theta})$ is a new parametrisation then $\mathbb{E}^*(\bm{\psi} \given Y = y) = T(\mathbb{E}^*(\bm{\theta} \given Y = y))$; this does not hold in general in a probabilistic setting \cite{druilhet2007invariant}, which makes the probabilistic MAP more subjective.

The considered framework allows for likelihoods defined as possibility functions, e.g.\ as the exponential of negative loss functions \cite{bissiri2016general} or negative energy functions \cite{ranzato2007efficient, ranzato2007unified}, which often model deterministic uncertainty. Moreover, since these functions naturally have a minimum equal to zero, the exponential of their negative counterparts have maximum one. We focus however on the case of a probabilistic likelihood in order to ease the comparison with standard Bayesian inference.

\subsection{Combining information}

One operation that is defined in general for possibility functions is the combination of information: if $\bm{\theta}$ and $\bm{\psi}$ are uncertain variables on $\Theta$ jointly described by the possibility function $f_{\bm{\theta},\bm{\psi}}$ on $\Theta \times \Theta$ and if we are given the information that $\bm{\theta}$ and $\bm{\psi}$ represent in fact the same unknown quantity then we can compute the conditional possibility function describing $\bm{\theta}$ (equiv.\ $\bm{\psi}$) given that $\bm{\theta} = \bm{\psi}$ as
\begin{equation}
\label{eq:combination}
f_{\bm{\theta}}(\theta \given \bm{\theta} = \bm{\psi}) = \dfrac{f_{\bm{\theta},\bm{\psi}}(\theta, \theta)}{\sup_{\theta' \in \Theta}f_{\bm{\theta},\bm{\psi}}(\theta', \theta')}, \qquad \theta \in \Theta.
\end{equation}
The normalising constant $c = \sup_{\theta \in \Theta}f_{\bm{\theta},\bm{\psi}}(\theta, \theta)$, which is assumed to be positive, corresponds to the credibility of the event $\bm{\theta} = \bm{\psi}$, i.e.\ to the credibility that $\bm{\theta}$ and $\bm{\psi}$ do represent the same unknown value; we will refer to $c$ as the \emph{consistency}. The conditioning used in \eqref{eq:combination} will be crucial for the proposed approach to robust inference but would lead to paradoxes \cite[Chapter 15.7]{jaynes2003probability} when applied in the context of probability theory. We will be particularly interested in the case where $\bm{\theta}$ and $\bm{\psi}$ are \emph{independently described}, that is $f_{\bm{\theta},\bm{\psi}}(\theta, \psi) = f_{\bm{\theta}}(\theta) f_{\bm{\psi}}(\psi)$ for any $\theta,\psi \in \Theta$. This notion of independence corresponds to the case where the available information about $\bm{\theta}$ is unrelated to the one about $\bm{\psi}$.

\subsection{Conjugate priors}

The concept of conjugate prior, which is central in Bayesian statistics, naturally extends to possibility functions \cite{houssineau2019elements} via \eqref{eq:Bayes_theorem}: if $f_{\bm{\theta}}$ and $f_{\bm{\theta}|Y}(\cdot \given y)$ take the same parametric form then $f_{\bm{\theta}}$ is said to be a conjugate prior (possibility function) for the likelihood $p_Y(y \given \cdot)$. In fact, each conjugate prior p.d.f.\ has an analogue as a possibility function up to shifts in the parameter set. For instance, for some $\mu \in \mathbb{R}^d$ and some $d \times d$ positive semidefinite matrix $P$,
\begin{equation*}
\overline{\text{N}}(\theta; \mu, P) = \exp\Big( -\dfrac{1}{2}(\theta - \mu)^{\intercal} P (\theta - \mu) \Big), \qquad \theta \in \mathbb{R}^d,
\end{equation*}
is the Gaussian/normal possibility function parametrised by its expected value $\mu$ and its precision matrix $P$ and is a conjugate prior for the multivariate normal likelihood $\text{N}(y; \theta, \Sigma)$ for any positive definite matrix $\Sigma$. The absence of normalising constant implies that $P$ does not have to be positive definite and, in fact, can be set to the $d \times d$ zero matrix $\bm{0}_{d,d}$ in which case $\overline{\text{N}}(\theta; \mu, \bm{0}_{d,d}) = \bm{1}$. Another example is the beta possibility function defined for some parameters $\alpha \geq 0$ and $\beta \geq 0$ as
\begin{equation*}
\overline{\text{B}}(\theta; \alpha, \beta) = \dfrac{(\alpha + \beta)^{\alpha + \beta}}{\alpha^{\alpha}\beta^{\beta}} \theta^{\alpha} (1 - \theta)^{\beta}, \qquad \theta \in [0,1],
\end{equation*}
with expected value $\alpha/(\alpha + \beta)$ and variance $\alpha\beta/(\alpha+\beta)^3$, which is a conjugate prior for Bernoulli and binomial likelihood like its probabilistic counterpart. Although the expected value matches between the probabilistic and possibilistic versions of the beta form, there is a shift by one unit in both parameters, as becomes apparent when noticing that $\overline{\text{B}}(0,0) = \text{B}(1,1) = \bm{1}$, with $\text{B}(\alpha,\beta)$ the beta p.d.f.\ with parameters $\alpha$ and $\beta$. The inverse-Wishart possibility function will also be introduced in the simulations.

As opposed to p.d.f.s, possibility functions are closed under power, i.e.\ $f_{\bm{\theta}}^{\gamma}$ remains a possibility function for any $\gamma \geq 0$. Also, if $\bm{\theta}$ is described by $f_{\bm{\theta}}$ and if $\bm{\psi}$ is described by $f_{\bm{\psi}} = f_{\bm{\theta}}^{\gamma}$ then $\mathbb{E}^*(\bm{\theta}) = \mathbb{E}^*(\bm{\psi})$ and $\mathbb{V}^*(\bm{\theta}) = \mathbb{V}^*(\bm{\psi})/\gamma$. Remarkably, conjugate prior families are also closed under power, with for instance $\overline{\text{N}}(\mu, P)^{\gamma} = \overline{\text{N}}(\mu, \gamma P)$ and $\overline{\text{B}}(\theta; \alpha, \beta)^{\gamma} = \overline{\text{B}}(\theta; \gamma\alpha, \gamma\beta)$. They also all contain the uninformative prior $\bm{1}$ which corresponds to the case where $\gamma = 0$. 

Although convenient, conjugate priors are not always applicable; in this case, synergies between probability and possibility theories can be leveraged to solve possibilistic inference problems using Markov chain Monte Carlo as in \cite{houssineau2021uncertainty} or sequential Monte Carlo as in \cite{ristic2018robust,ristic2020target}. This is similar to the use of random exploration in optimisation problems, but with the definition of the underlying p.d.f.s being directly motivated, e.g.\ by seeing o.p.m.s as upper bounds for p.d.f.s.

\section{Robust inference}
\label{sec:robust_inference}

\subsection{From evidence to consistency}

The evidence $e(y)$ behaves differently from the evidence $e'(y) = \int p_Y(y \given \theta)\pi(\theta) \mathrm{d} \theta$ that would be obtained in the standard approach with a prior probability distribution $\pi$ on $\Theta$. Considering for instance the case where $\Theta = \mathbb{R}$, it holds that one of the prior possibility functions $f_{\bm{\theta}}$ maximising $e(y)$ is the uninformative prior $\bm{1}$; this is in stark contrast with $e'(y)$ which tends to $0$ if $\pi(\theta) \propto \bm{1}_B(\theta)$ with $B = [-s,s]$ and $s \to \infty$. Improper prior distributions cannot be used to evaluate the evidence as $e'(y)$ would essentially be arbitrary in this case. While the behaviour of $e'(y)$ is ideal for model selection where the fitness of the prior is key, the evidence $e(y)$ appears to be suitable for evaluating the consistency between the observation and the prior. However, the value of $e(y)$ depends on the choice of the reference measure when defining the likelihood $p_Y(y \given \theta)$, which introduces some arbitrariness. To circumvent this dependence, we first characterise the information in the likelihood via the posterior possibility function
\begin{equation}
\label{eq:likelihood_info}
f_{\bm{\psi}|Y}(\psi \given y) = \dfrac{p_Y(y \given \psi)}{\sup_{\psi' \in \Theta} p_Y(y \given \psi')}, \qquad \psi \in \Theta,
\end{equation}
where we have assumed that the likelihood $p_Y(y \given \cdot)$ is bounded. Indeed, the possibility function $f_{\bm{\psi}|Y}(\cdot \given y)$, which is related to the frequentist notion of likelihood ratio test, only contains information stemming from the likelihood and from the received observation~$y$. On the other hand, we have some prior information described by $f_{\bm{\theta}}$, and we assume as is usual that the information encoded in $f_{\bm{\theta}}$ is unrelated to either the likelihood or the observation. It follows that $\bm{\theta}$ and  $\bm{\psi}|Y$ are independently described and the underlying information can be combined as
\begin{equation*}
f_{\bm{\theta}}(\theta \given \bm{\theta} = \bm{\psi}, Y = y) = \dfrac{f_{\bm{\psi}|Y}(\theta \given y) f_{\bm{\theta}}(\theta)}{\sup_{\theta' \in \Theta} f_{\bm{\psi}|Y}(\theta' \given y) f_{\bm{\theta}}(\theta')} = f_{\bm{\theta}|Y}(\theta \given y), \qquad \theta \in \Theta.
\end{equation*}
Although the resulting possibility function $f_{\bm{\theta}}(\cdot \given \bm{\theta} = \bm{\psi}, Y = y)$ is equal to the posterior possibility function $f_{\bm{\theta}|Y}(\cdot \given y)$, the corresponding consistency $c(y) = \sup_{\theta \in \Theta} f_{\bm{\psi}|Y}(\theta \given y) f_{\bm{\theta}}(\theta)$ differs from $e(y)$. The advantages of using $c(y)$ rather than $e(y)$ are as follows: the scalar $c(y)$ does not depend on the choice of reference measure when defining the likelihood and it is \emph{calibrated}, i.e.\ $c(y) \in [0,1]$, with $c(y) = 1$ being achieved when $f_{\bm{\theta}} = \bm{1}$. Henceforth, for the sake of simplicity, we will write $L(\theta \given y)$ instead of $f_{\bm{\psi}|Y}(\theta \given y)$ for the information in the likelihood and we will suppress the explicit conditioning on the observations in the indices, e.g.\ we will write $f_{\bm{\theta}}(\cdot \given y)$ instead of $f_{\bm{\theta}|Y}(\cdot \given y)$ for the posterior possibility function given $Y = y$.

\subsection{Discounting the information in the likelihood}
\label{sec:original}

Discounting or tempering some or all of the components in statistical inference is a standard procedure, from the classical composite likelihood \cite{varin2011overview} to the more recent variational tempering \cite{mandt2016variational}, and it is particularly meaningful for possibility functions which are closed under powers. The objective in this section is to introduce a suitable discounting mechanism for the information in the likelihood in order to gain robustness properties. We first consider an example in order to better understand how the consistency behaves, especially w.r.t.\ the dimension of the parameter set.

\begin{example}
\label{ex:dimension}
Consider a likelihood of the form $p_Y(y \given \theta) = p_{Y'}( y_1 \given \theta_1 ) \dots p_{Y'}( y_d \given \theta_d )$ with observation $y = (y_1,\dots,y_d)$ and parameter $\theta = (\theta_1,\dots,\theta_d) \in \Theta^d$. This corresponds to the case where $d$ conditionally i.i.d.\ observations are generated based on $d$ different parameters. In this case, the consistency of the observation $y$ w.r.t.\ an independent prior $f_{\bm{\theta}}(\theta) = f_{\bm{\theta}'}(\theta_1) \dots f_{\bm{\theta}'}(\theta_d)$ takes the form
\begin{equation*}
c(y) = \prod_{i=1}^d \bigg[ \sup_{\theta \in \Theta} L'(\theta_i \given y_i) f_{\bm{\theta}'}(\theta_i) \bigg] = \prod_{i=1}^d c'(y_i),
\end{equation*}
where $L'(\theta_i \given y_i) \propto p_{Y'}(y_i \given \theta_i)$ is the information in the $i$-th component of the likelihood. The expression of the consistency $c(y)$ as a product of the component-wise consistencies $c'(y_i) \in [0,1]$, $i \in \{1,\dots,d\}$ shows that $c(y)$ will tend to decrease as the dimension $d$ increases. We could however consider the geometric average $\sqrt[d]{c(y)}$ 
to compensate for this dependence on the dimension.
\end{example}

A candidate for a robust posterior possibility function follows from the observation made in Example~\ref{ex:dimension} as
\begin{equation}
\label{eq:robust_posterior}
f^{\mathrm{r}}_{\bm{\theta}}(\theta \given y) \doteq \dfrac{L(\theta \given y)^{\gamma} f_{\bm{\theta}}(\theta)}{\sup_{\theta' \in \Theta} L(\theta' \given y)^{\gamma} f_{\bm{\theta}}(\theta')}, \qquad \theta \in \Theta,
\end{equation}
where the discount $\gamma$ is equal to $\sqrt[d]{c(y)}$, 
with the consistency $c(y)$ defined as
\begin{equation*}
c(y) = \sup_{\theta \in \Theta} L(\theta \given y) f_{\bm{\theta}}(\theta).
\end{equation*}
If the prior and observation are fully consistent, i.e.\ $c(y) = 1$, then there is no discount and the robust posterior $f^{\mathrm{r}}_{\bm{\theta}}(\cdot \given y)$ is equal to the standard posterior $f_{\bm{\theta}}(\cdot \given y)$. On the other hand, it is no longer necessary to assume that $c(y) > 0$ since $c(y) = 0$ leads to a well-defined robust posterior which is in fact equal to the prior $f_{\bm{\theta}}$. The idea of discounting the likelihood to obtain robustness has been explored in the standard Bayesian context in \cite{miller2019robust} and in a generalised Bayesian context in \cite{knoblauch2018doubly,boustati2020generalised} although, in both cases, the value of the discount has to be separately optimised in general. In the context of prior-data checks \cite{bayarri2007bayesian,nott2020checking,nott2021using}, one would generally apply the discount $\gamma$ to the prior $f_{\bm{\theta}}$, which would be equally well-defined in our formulation.



We now consider the case where several observations $y_1,\dots,y_n$ are available as realisations of i.i.d.\ copies of $Y$. One important property of posterior probability distributions that extends to posterior possibility functions is that the way in which the observations are taken into account does not affect the overall posterior: updating with $(y_1,y_2)$ is the same as updating with $y_1$ and $y_2$ separately. This property no longer holds with the robust posterior since the consistency of $y_1$ and $y_2$ will be jointly assessed when updating with $(y_1,y_2)$; for instance, if $y_1$ is an outlier then both $y_1$ and $y_2$ will be mostly ignored when jointly updating with $(y_1,y_2)$. It follows that independent observations should be taken into account one by one in order to retain as much information as possible. However, this statement only applies to situations where each observation is sufficiently informative to yield a meaningful assessment of the consistency. This is particularly crucial for parameters that are not observed directly, unlike most location parameters for instance. This aspect is illustrated in the following example.

\begin{example}
\label{ex:Bernoulli}
Consider the observations $y_1,\dots,y_n \in \{0,1\}$ corresponding to Bernoulli experiments with unknown probability of success $\theta \in \Theta = [0,1]$ together with a prior beta possibility function. In this case, the information in a single observation $y_i$, $i \in \{1,\dots,n\}$, is so limited that the ensuing assessment of the consistency bears little meaning. For instance, if we have two differently biased coins, then one cannot determine which coin is being flipped by looking at the data from a single flip. If $n$ is sufficiently large, one can however consider the binomial likelihood corresponding to the observation $y = y_1 + \dots + y_n$. Taking $n$ large is not an issue as long as the observations $y_1,\dots,y_n$ are either all inliers or all outliers.
\end{example}

\section{Limitations and extensions}
\label{sec:limitations}

\subsection{Ordering}

A property of standard Bayesian inference that does not extend to the proposed approach to robust inference is the independence w.r.t.\ the order in which the observations are taken into account. This is not an issue in models where there is a natural order between observations such as time series, however, it is an aspect to be considered in the absence of such an order. We will show in the simulations that the ordering has little impact on the results obtained with the proposed approach when assuming that the first observation is an inlier. Regardless of the existence of a natural order between observations, a generalisation of the proposed approach would be needed in order to allow for the first observation to be an outlier. Indeed, in this case, the information in both the likelihood and the prior should be allowed to be discounted in the presence of evidence against one or the other. In particular, prior and likelihood should be similarly discounted if the information in the prior is of the same order as the one in the likelihood; this will be the topic of future work.

\subsection{Information loss}
\label{sec:alternative}

The two main advantages of the approach based on \eqref{eq:robust_posterior} are the absence of parameters in the way outliers are dealt with as well as the fact that conjugate prior families can still be used. Yet, the fact that the likelihood for most observations will be discounted to some extent means that there is an information loss between the robust posterior defined in \eqref{eq:robust_posterior} and the standard posterior based on inliers only. To reduce this information loss one can introduce a threshold $\tau \in (0,1]$ such that no discount is applied when $\gamma > \tau$ and the usual discount of gamma is applied otherwise. The corresponding posterior possibility function is
\begin{equation}
\label{eq:threshold_posterior}
f^{(\tau)}_{\bm{\theta}}(\theta \given y) \doteq \dfrac{L(\theta \given y)^{\rho(\tau)} f_{\bm{\theta}}(\theta)}{\sup_{\theta' \in \Theta} L(\theta' \given y)^{\rho(\tau)} f_{\bm{\theta}}(\theta')},
\end{equation}
where $\rho(\tau)$ equals $\gamma$ if $\gamma \leq \tau$ and $1$ otherwise. The value of the parameter $\tau$ does not depend strongly on the considered statistical model, and should be chosen so as to separate the values that $\sqrt[d]{c(y)}$ 
takes for inliers from the ones it takes for outliers. It would be possible to consider $\rho(\tau) = 0$ when $\gamma \leq \tau$, however, this would prevent the algorithm from recovering from early outliers that might have been assimilated due to the lack of information.

\subsection{Unbounded likelihood}
\label{sec:unbounded_likelihood}

The information in the likelihood $L(\psi \given y) = f_{\bm{\psi}}(\psi \given y) \propto p_Y(y \given \psi)$, $\psi \in \Theta$, might be undefined if the likelihood is unbounded. In this situation, one can make the prior possibility function $f_{\bm{\psi}}$ weakly informative in order to ensure that $\psi \mapsto p_Y(y \given \psi) f_{\bm{\psi}}(\psi)$ is bounded. The prior $f_{\bm{\psi}}$ should however be independent from $f_{\bm{\theta}}$ so as to avoid any redundancy when combining the latter with $L(\cdot \given y)$. Yet, when multiple observations are available, the same prior $f_{\bm{\psi}}$ should not be used for each observation as this will artificially inflate the available information. At iteration $t \in \{1,\dots,T\}$, one can split the information in the prior $f_{\bm{\theta}}(\cdot \given y_1, \dots, y_{t-1})$ into two independent bits of information as follows
\begin{equation*}
f^{\mathrm{r}}_{\bm{\theta}}(\theta \given y_1, \dots, y_t) \doteq \dfrac{L_t(\theta \given y_t)^{\gamma_t} f^{\mathrm{r}}_{\bm{\theta}}(\theta \given y_1, \dots, y_{t-1})^{\omega}}{\sup_{\theta' \in \Theta} L_t(\theta' \given y_t)^{\gamma_t} f^{\mathrm{r}}_{\bm{\theta}}(\theta' \given y_1, \dots, y_{t-1})^{\omega}}
\end{equation*}
for some $\omega \in (0,1)$, where $\gamma_t = \sup_{\theta \in \Theta} p_Y(y_t \given \theta) f^{\mathrm{r}}_{\bm{\theta}}(\theta \given y_1, \dots, y_{t-1})^{1-\omega}$ and
\begin{equation*}
L_t(\theta \given y_t) = \gamma_t^{-1} p_Y(y_t \given \theta) f^{\mathrm{r}}_{\bm{\theta}}(\theta \given y_1, \dots, y_{t-1})^{1-\omega}.
\end{equation*}
From the viewpoint of possibility theory, $f_{\bm{\theta}}^{\omega}$ and $f_{\bm{\theta}}^{1-\omega}$ are independent for any possibility function $f_{\bm{\theta}}$ and any $\omega \in [0,1]$, so that there is no unwanted reuse of information in the posterior $f^{\mathrm{r}}_{\bm{\theta}}(\cdot \given y_1, \dots, y_t)$. The interpretation of this recursion is as follows: $100(1-\omega)\%$ of the posterior information given by $y_1, \dots, y_{t-1}$ is used to ensure that $L_t(\theta \given y_t)$ is well defined and the rest is used as a prior. The parameter $\omega$ should be kept close to $1$ since the amount of information invested in $L_t(\theta \given y_t)$ might be lost if the observation $y_t$ is an outlier. A numerical example including an unbounded likelihood can be found in Appendix~\ref{sec:s_unbounded}.

\subsection{Likelihood with observation-dependent support}
\label{sec:observation-dependent_support}

The reliance of the proposed approach on discounting means that the likelihood cannot be tempered if it is equal to $0$ on parts of the parameter space. This might be problematic as soon as the likelihood has an observation-dependent support, such as with the uniform likelihood $p_Y(y \given \theta) = \theta^{-1}\bm{1}_{[y, \infty)}(\theta)$. Indeed, in this case, the likelihood will overwrite the prior regardless of the consistency. One solution is to change the likelihood to make sure that $p_Y(y \given \cdot)$ has the same support for all $y$, e.g.\ by setting exponential decays where the likelihood was originally equal to $0$. An example of this is given in Appendix~\ref{sec:s_observation-dependent}. The main shortcoming of such an approach is that additional parameters have to be introduced in order to define how fast the likelihood will tend to $0$. The same limitation would apply to standard probabilistic methods; however, it is possible to work in the space of probability distributions instead \cite{minsker2017robust, cheriefabdellatif2020mmd}, in which case a likelihood of $0$ can be overcome by defining a suitable notion of distance, e.g.\ the Wasserstein metric.

\section{Simulations}
\label{sec:simulations}

We present three challenging applications for our method, the first and second on simulated data and the third one on real data. These applications show the generality of our approach as well as the more general strengths of possibilistic inference.

\subsection{Extended feature estimation}
\label{sec:extended_feature}

We consider the estimation of the location and extent of a Gaussian-shaped feature in dimension $d$. The data is simulated so that the performance of the approach can be better assessed. The extent of the feature is modelled by a positive definite matrix $\Sigma$, sampled at random from the inverse-Wishart distribution $\text{IW}(I_d,d\sqrt{d})$, with $I_d$ the identity matrix of dimension $d$. There are $T \in \mathbb{N}$ iterations and, at each iteration $t \in \{1,\dots,T\}$, we receive $n$ observations $y_{t,i}$, $i \in \{1,\dots,n\}$, which are sampled independently from a normal distribution with mean zero and covariance $\Sigma$. We define $\mathcal{Y}_t$ as the set $\{y_{t,1},\dots,y_{t,n}\}$. From an inference viewpoint, the possibility function
\begin{equation*}
\overline{\text{NIW}}(\mu, \Sigma; \mu_0, \lambda, \Psi, \nu) = \overline{\text{N}}( \mu; \mu_0, \lambda \Sigma^{-1} ) \overline{\text{IW}}(\Sigma; \Psi, \nu)
\end{equation*}
is a conjugate prior for the likelihood $\prod_{i=1}^n \text{N}(y_{t,i}; \mu, \Sigma)$, where $\overline{\text{IW}}(\Psi, \nu)$ is the inverse Wishart possibility function with $\Psi \in \mathbb{R}^{d \times d}$ positive semi-definite and $\nu \geq 0$, which is defined as
\begin{equation*}
\overline{\text{IW}}(\Sigma; \Psi, \nu) = \bigg[ \dfrac{|\Psi|}{|\nu \Sigma|} \bigg]^{\nu/2} \exp\bigg( -\dfrac{1}{2} \big( \tr(\Psi\Sigma^{-1}) - d\nu \big) \bigg),
\end{equation*}
with $|\cdot|$ and $\tr(\cdot)$ respectively denoting the determinant and the trace. The matrix $\Psi$ is allowed not to be positive definite when $\nu = 0$. Outliers are generated with probability $\epsilon$ and are sampled from the same data generating process as the inliers but with an inflated covariance of $\alpha \Sigma$ with $\alpha > 1$.

We model the unknown mean $\mu$ and covariance $\Sigma$ by the respective uncertain variables $\bm{\mu}$ and $\bm{\Sigma}$ and define $\bm{\theta} = (\bm{\mu}, \bm{\Sigma})$. We define the prior possibility function $f_{\bm{\theta}}$ as uninformative, i.e.\ $f_{\bm{\theta}} = \bm{1}$, which corresponds to the $\overline{\text{NIW}}(\mu_0, \lambda_0, \Psi_0, \nu_0)$ possibility function when $\lambda_0 = \nu_0 = 0$ and $\Psi_0 = \bm{0}_{d,d}$. 
The updated parameters in the discounted normal inverse-Wishart model are
\begin{align*}
\mu_t & = \dfrac{\lambda_{t-1} \mu_{t-1} + \gamma_t n \bar{y}_t}{\lambda_{t-1} + \gamma_t n} \\
\lambda_t & = \lambda_{t-1} + \gamma_t n \\
\Psi_t & = \Psi_{t-1} + \gamma_t S_t + \dfrac{\lambda_{t-1}\gamma_t n}{\lambda_{t-1} + \gamma_t n} (\mu_{t-1} - \bar{y}_t)^{\intercal}(\mu_{t-1} - \bar{y}_t)) \\ 
\nu_t & = \nu_{t-1} + \gamma_t n,
\end{align*}
where $\gamma_t$ is the discount, where $S_t = \sum_{i=1}^n (y_i - \bar{y}_t)^{\intercal}(y_i - \bar{y}_t)$, and where $\bar{y}_t$ is the mean of the observations $y_{t,1},\dots,y_{t,n}$. The non-discounted parameters used in the expression of $c_t$ in Section~\ref{sec:extended_feature} can be easily recovered by setting $\gamma_t = 1$.

The consistency $c_t \doteq c_t(\mathcal{Y}_t)$ verifies
\begin{equation*}
c_t^2 = \bigg|\dfrac{S_t}{n}\bigg|^n \bigg|\dfrac{\Psi_{t-1}}{\nu_{t-1}}\bigg|^{\nu_{t-1}} \bigg|\dfrac{\Psi_t}{\nu_t}\bigg|^{-\nu_t},
\end{equation*}
where $S_t = \sum_{i=1}^n (y_i - \bar{y}_t)^{\intercal}(y_i - \bar{y}_t)$, with $\bar{y}_t$ the mean of the observations $y_{t,1},\dots,y_{t,n}$.

We consider in particular the case where the dimension is $d=10$ and where there are $n = 25$ observations at each iteration $t \in \{1,\dots,T\}$ with $T = 500$. The probability of outliers is $\epsilon = 0.02$ and we consider an inflation of the variance by a coefficient $\alpha = 3$. The discount at iteration $t$ is defined as $\gamma_t = \sqrt[d']{c_t}$ 
with $d' = d + d(d+1)/2$ the effective dimension of $\bm{\theta}$ which follows from the fact that $\bm{\Sigma}$ is positive definite. We compare the original method of Section~\ref{sec:original} with the alternative approach described in Section~\ref{sec:alternative}, the latter being implemented with a threshold $\tau = 0.25$. The estimation error is defined as the Frobenius norm $\| \Sigma - \mathbb{E}^*(\bm{\Sigma}|\mathcal{Y}_{1:T}) \|_{\mathrm{F}}$, with $\mathbb{E}^*(\bm{\Sigma}|\mathcal{Y}_{1:T}) = \Psi_T/\nu_T$ the posterior expected value of $\bm{\Sigma}$.

\begin{figure}
\begin{subfigure}[b]{0.48\textwidth}
\includegraphics[trim=30pt 0pt 55pt 20pt,clip,width=\textwidth]{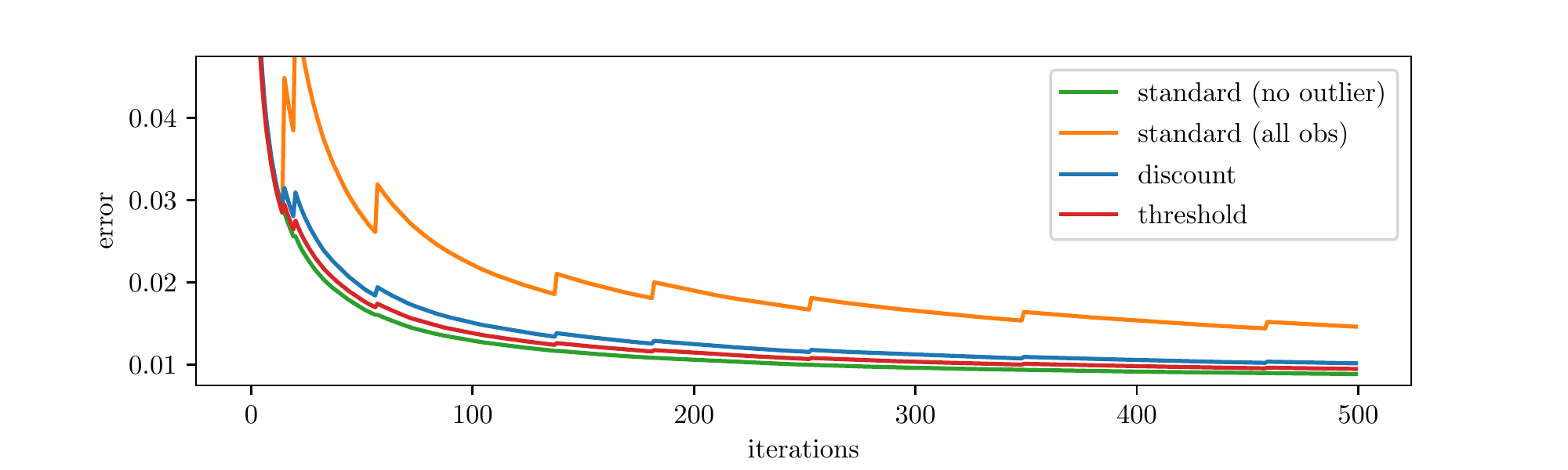}
\caption{Estimation error in Frobenius norm, averaged over $1000$ repeats.}
\label{fig:feature_error}
\end{subfigure}\quad
\begin{subfigure}[b]{0.48\textwidth}
\includegraphics[trim=35pt 0pt 55pt 20pt,clip,width=\textwidth]{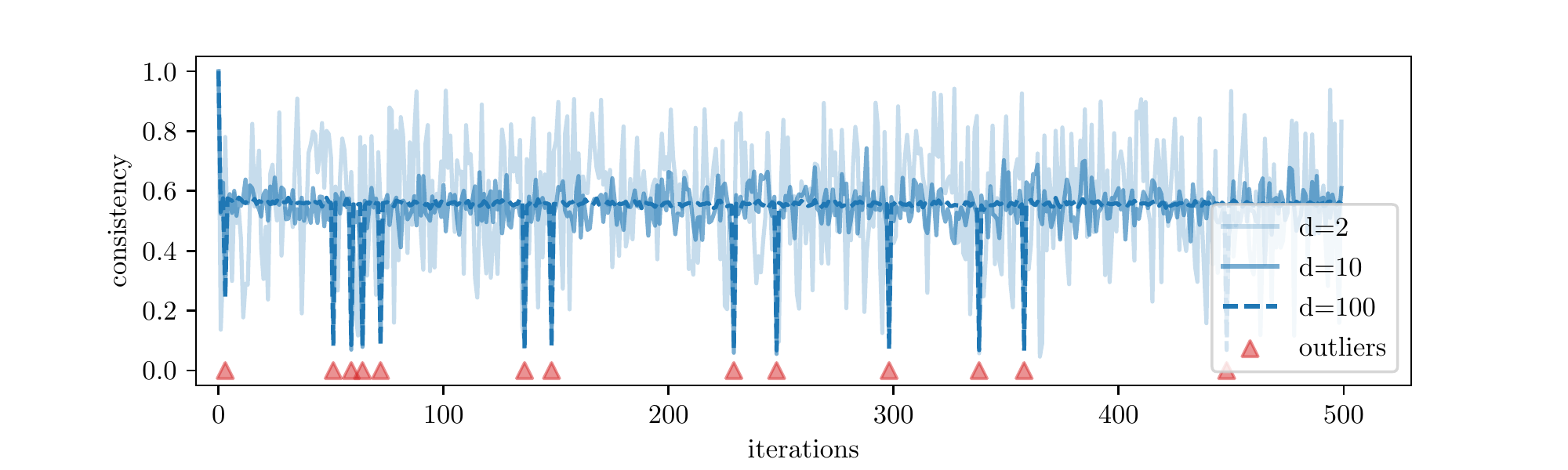}
\caption{Consistency per iteration, for $3$ different dimensions and sample sizes.}
\label{fig:feature_dimension}
\end{subfigure}
\caption{Results for the extended feature estimation.}
\end{figure}

We first show that the results depend weakly on the arbitrary order in which the observations are considered. For this purpose, we fix a realisation of $\Sigma$ and of the observations $\mathcal{Y}_{1:T}$ and consider $1000$ random permutations of the indices $\{1,\dots,T\}$ with the constraint that the observation at the first iteration after permutation must not be an outlier. The resulting standard deviation of the error is less than $2 \times 10^{-4}$ for both methods, which confirms the weak dependence of the error on the ordering of the observations. Next, we compute the evolution of the error as a function of the iteration $t \in \{1,\dots,T\}$ in $4$ different cases:
\begin{enumerate*}[label=\roman*)]
\item standard Bayesian inference without outliers,
\item standard Bayesian inference with all observations,
\item the original method, and
\item the alternative method with a threshold $\tau = 0.25$.
\end{enumerate*}
The results displayed in Figure~\ref{fig:feature_error} show that the use of the consistency as a discount in the likelihood, as advocated in Section~\ref{sec:original}, allows for largely compensating for the presence of outliers when compared to the standard Bayesian recursion. The use of a threshold allows for gaining further accuracy at the cost of adding a tuning parameter. Figure~\ref{fig:feature_dimension} illustrates the effect of the dimension on the consistency, with a sample size of $n = 5d/2$; it shows that, with such a linear dependence between dimension and sample size, the respective consistencies of inliers and outliers tend to stabilise when the dimension increases. The characterisation of this tendency is the topic of future work.

\subsection{Kalman filtering}
\label{sec:s_Kalman_filtering}

We now consider a truly dynamical problem where preserving the closed-form recursion of existing solutions is particularly crucial. One of the most important tools for dynamical systems is the Kalman filter which applies in the linear-Gaussian case. The Kalman filter can be shown to hold with an identical recursion in the context of possibility theory \cite{houssineau2018smoothing} and the objective of this section is to illustrate how it can easily be made more robust with the proposed approach. There has been a lot of research on robust versions of the Kalman filter, see e.g.\ \cite{xie1994robust,simon2006optimal,gandhi2009robust} as well as more generally for online inference with hidden Markov models \cite{maiz2012particle,boustati2020generalised}. For instance, the method introduced in \cite{gandhi2009robust} requires the use of a pre-whitening method based on the sample median and MAD, followed by an application of the iteratively-reweighted least squares algorithm; in comparison the approach we describe is a simple modification of the Kalman filter with negligible effects on the computational time.

Specifically, we consider a scenario where the state follows a nearly-constant velocity model over $T$ time steps of duration $\Delta$, i.e.\ $\bm{\theta}_t$ is described by $\overline{\text{N}}(F \theta_{t-1}, Q)$ given that $\bm{\theta}_{t-1} = \theta_{t-1}$, with $F$ the transition matrix and $Q$ the covariance matrix of the dynamical noise. We assume that the observation $y_t$ is the realisation of a normal random variable with mean $H \theta_t$, with $H$ the observation matrix, and with covariance matrix $R$. In this context, we have $L(\theta_t \given y_t) = \overline{\text{N}}(\theta_t; y_t, R)$ and the consistency is
\[
c_t(y_t) = \overline{\text{N}}(y_t; H \hat{\theta}_t, H \Sigma_t H^{\intercal} + R),
\]
with $\mu_t$ the predicted expected value at time $t$ and $\Sigma_t$ the corresponding covariance matrix. This result is the direct analogue of the marginal likelihood in a probabilistic context.

\begin{figure}
\centering
\begin{subfigure}[b]{0.48\textwidth}
\includegraphics[trim=24pt 5pt 55pt 40pt,clip,width=\textwidth]{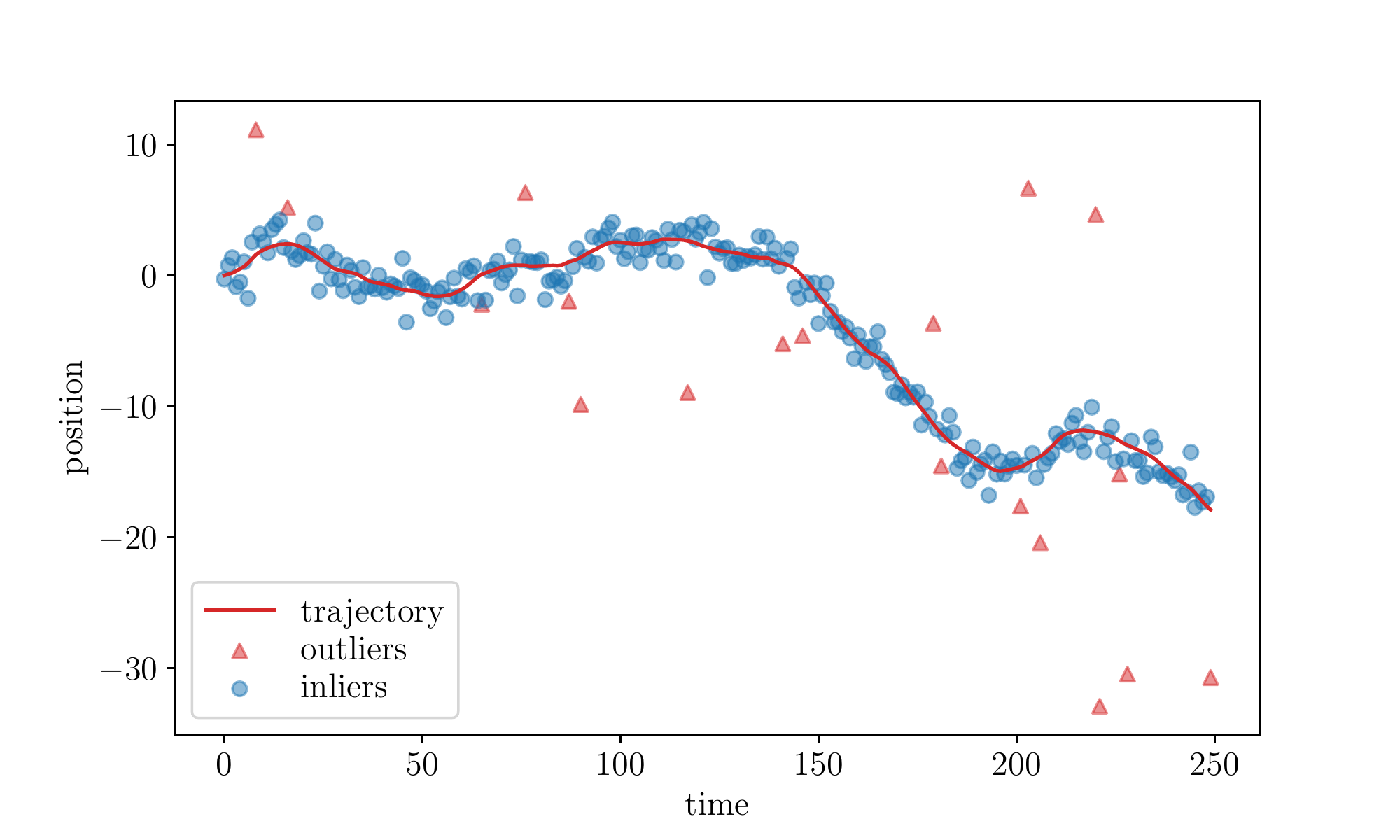}
\caption{Realisation of the position (first component of the state) and observation processes.}
\label{fig:constant-velocity_model}
\end{subfigure}\quad
\begin{subfigure}[b]{0.48\textwidth}
\includegraphics[trim=24pt 5pt 55pt 40pt,clip,width=\textwidth]{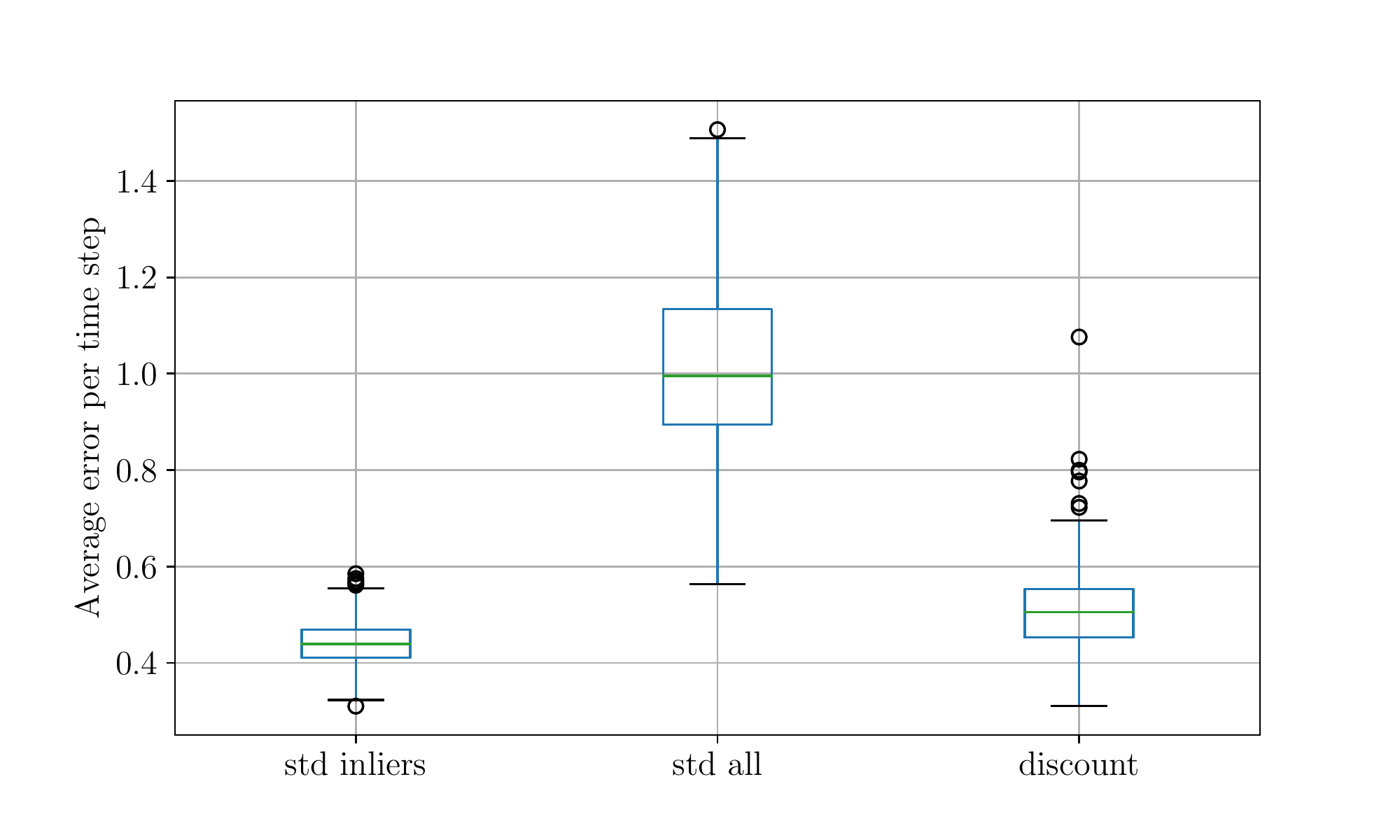}
\caption{Box plots of the absolute error averaged over the $T$ time steps for $500$ repeats.}
\label{fig:constant-velocity_results}
\end{subfigure}
\caption{Generated data and results for the model of Section~\ref{sec:s_Kalman_filtering}}
\end{figure}

We consider in particular a nearly-constant velocity model over $T = 250$ time steps of duration $\Delta = 1$, which is defined via
\[
F = \begin{bmatrix}
1 &  \Delta \\
0 & 1
\end{bmatrix}
\qquad \text{and} \qquad
Q = \sigma_{\mathrm{a}}^2 \begin{bmatrix}
\Delta^4/4 & \Delta^3/2 \\
\Delta^3/2 & \Delta^2 \\
\end{bmatrix},
\]
where $\sigma_{\mathrm{a}} = 0.05$ is the standard deviation of the acceleration noise. The observation matrix is $H = \begin{bmatrix} 1 & 0 \end{bmatrix}$ and we consider $R = 1$. The true state is initialised at $\theta^*_1 = \begin{bmatrix} 0 & 0.1 \end{bmatrix}^{\intercal}$ and is propagated according to $\theta^*_t = F\theta^*_{t-1} + u_t$ with $u_t \sim \text{N}(0,Q)$ independently from all other variables. Outliers are generated with probability $\epsilon = 0.1$ and are sampled, at time $t$, from a normal distribution with mean $H\theta^*_t$ and standard deviation $10$. A realisation of the state and observation processes for the considered scenario is displayed in Figure~\ref{fig:constant-velocity_model} as an illustration. The performance of the proposed approach is shown in Figure~\ref{fig:constant-velocity_results} which includes box plots of the time-averaged absolute error for $500$ repeats for each of the considered methods: ``std inliers'' for the standard Bayesian approach with inliers only, ``std all'' for the standard Bayesian approach with all observations and ``discount'' for the proposed approach.

\subsection{Change-point detection}
\label{sec:changepoint}

We formulate Bayesian change-point detection \cite{adams2007bayesian, fearnhead2007line, knoblauch2018doubly} in the context of possibility theory as follows: a sequence of observations $y_1,\dots,y_T$ is divided into intervals of unknown run lengths, which induces a partition $\Pi$ of $\{1,\dots,T\}$. The objective is to detect the interfaces between these intervals which are referred to as change points. For the sake of simplicity, we assume that the observations within one subset $S$ of the partition are conditionally i.i.d.\ according to $p_Y(\cdot \given \theta_S)$, $\theta_S \in \Theta$, with $\theta_S \neq \theta_{S'}$ for any $S, S' \in \Pi$ such that $S \neq S'$. We introduce an integer-valued uncertain variable $\bm{r}_t$ modelling the unknown run-length of the current interval as well as a $\Theta$-valued uncertain variable $\bm{\theta}$ modelling the unknown parameter for the current interval. The run-length $\bm{r}_t$ is described by the possibility function $f_{\bm{r}_t|\bm{r}_{t-1}}$ defined as
\begin{equation*}
f_{\bm{r}_t|\bm{r}_{t-1}}(r \given r') =
\begin{cases}
h(r'+1) & \text{if } r = 0 \\
1 & \text{if } r = r' + 1 \\
0 & \text{otherwise},
\end{cases}
\end{equation*}
where $h(r) \in [0,1]$ models the credibility that the total run-length of the current interval is equal to $r$. Since possibility functions can be seen as upper bounds for p.d.f.s (via o.p.m.s), $h(r)$ can be seen as the maximum subjective probability for a change point to occur after $r$ iterations. Although it would be possible to set $h(r)$ to $1$ for all $r$, this would model the fact that there is no prior information on durations between change points, which would make the detection of the latter impossible; indeed, in this situation, the most likely outcome would be that a change point occurs at every iteration. Yet, one can model the limited information about interval lengths by setting $h(r)$ to a sufficient large value. The predicted possibility function as iteration $t$ can now be expressed as
\begin{equation*}
f_{\bm{r}_t, \bm{\theta}}(r, \theta \given y_{1:t-1}) \propto
\begin{cases}
\displaystyle \max_{r' > 0} f_{\bm{r}_t|\bm{r}_{t-1}}(r \given r') \hat{f}_{\bm{r}_{t-1}, \bm{\theta}}(r', \theta \given y_{1:t-1}) & \text{if } r = 0 \\
f_{\bm{r}_t|\bm{r}_{t-1}}(r \given r-1) \hat{f}_{\bm{r}_{t-1}, \bm{\theta}}(r-1, \theta \given y_{1:t-1}) & \text{otherwise},
\end{cases}
\end{equation*}
where $y_{t':t}$ stands for the sequence $(y_{t'},\dots,y_t)$ when $t' \geq t$ and for the empty sequence when $t' > t$ and where $\hat{f}_{\bm{r}_{t-1}, \bm{\theta}}(\cdot \given y_{1:t-1})$ is the posterior at iteration $t-1$. The predicted possibility function describing $\bm{\theta}$ given $\bm{r}_t$ satisfies
\begin{equation*}
f_{\bm{\theta}|\bm{r}_t}(\theta \given r, y_{1:t-1}) = f_{\bm{\theta}}(\theta \given y_{t-r:t-1}),
\end{equation*}
that is, $\bm{\theta}$ only depends on the last $\bm{r}_t$ observations and does not otherwise depend on $\bm{r}_t$; in particular, $f_{\bm{\theta}|\bm{r}_t}(\theta \given 0, y_{1:t})$ is equal to the prior possibility function $f_{\bm{\theta}}$ describing $\bm{\theta}$. The posterior possibility function at iteration $t$ is then characterised by
\begin{equation*}
\hat{f}_{\bm{r}_t, \bm{\theta}}(r, \theta \given y_{1:t}) \propto p_Y(y_t \given \theta) f_{\bm{r}_t, \bm{\theta}}(r, \theta \given y_{1:t-1}).
\end{equation*}
As in the probabilistic algorithmic solutions to this problem, the tail of the posterior $f_{\bm{r}_t}(\cdot \given y_{1:t})$ has to be truncated in practice in order to control the computational cost; however, this truncation does not affect the possibility of non-truncated values of $\bm{r}_t$ since no normalisation is needed.

Change-point detection in the presence of outliers is particularly challenging in a filtering setting since only future observations can help distinguish between the two. In order to devise a robust solution to the change-point detection problem, we introduce the consistency $c_t(y_t)$ at iteration $t$ as
\begin{equation*}
c_t(y_t) = \sup_{(\theta,r) \in \Theta \times \mathbb{N}_0} p_Y(y_t \given \theta) f_{\bm{r}_t, \bm{\theta}}(r, \theta \given y_{1:t-1}).
\end{equation*}
Although the likelihood only depends on $\bm{\theta}$, the consistency $c_t(y_t)$ also take into account the run-length $\bm{r}_t$ via the conditional $f_{\bm{\theta}|\bm{r}_t}(\cdot \given r, y_{1:t-1})$; e.g., if a change most likely just happened ($\mathbb{E}^*(\bm{r}_t) = 0$) and if $f_{\bm{\theta}}$ is uninformative then $c_t(y_t) = 1$. The robust posterior can then be defined recursively as
\begin{equation*}
\hat{f}^{\mathrm{r}}_{\bm{r}_t, \bm{\theta}}(r, \theta \given y_{1:t}) = \dfrac{L(\theta \given y_t)^{\gamma_t} f^{\mathrm{r}}_{\bm{r}_t,\bm{\theta}}(r, \theta \given y_{1:t-1})}{\sup_{(\theta', r') \in \Theta \times \mathbb{N}_0} L(\theta' \given y_t)^{\gamma_t} f^{\mathrm{r}}_{\bm{r}_t,\bm{\theta}}(r', \theta' \given y_{1:t-1})},
\end{equation*}
where $L(\theta \given y_t) \propto p_Y(y_t \given \theta)$ and where $\gamma_t = \sqrt{c_t(y_t)}$ since this model is two-dimensional.

\begin{figure}
\begin{subfigure}[b]{\textwidth}
\includegraphics[trim=100pt 40pt 110pt 40pt,clip,width=\textwidth]{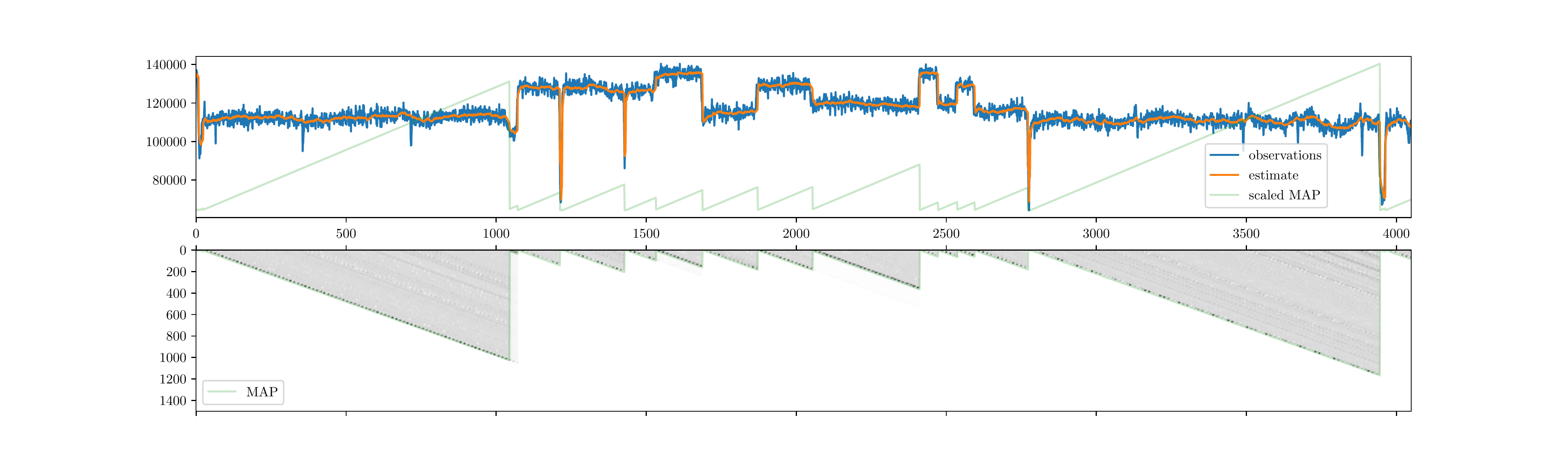}
\caption{Results on the well-log data with (bottom panel) a representation of the possibility function $\sqrt[4]{f_{\bm{r}_t}}$ 
as well as the MAP in terms of run-length against time and (top panel) observations and estimate together with a rescaled version of the MAP in terms of run-length. The possibility function $f_{\bm{r}_t}$ is rescaled for visibility purposes.}
\label{fig:welllog_results}
\end{subfigure}
\begin{subfigure}[b]{\textwidth}
\includegraphics[trim=100pt 0pt 110pt 0pt,clip,width=\textwidth]{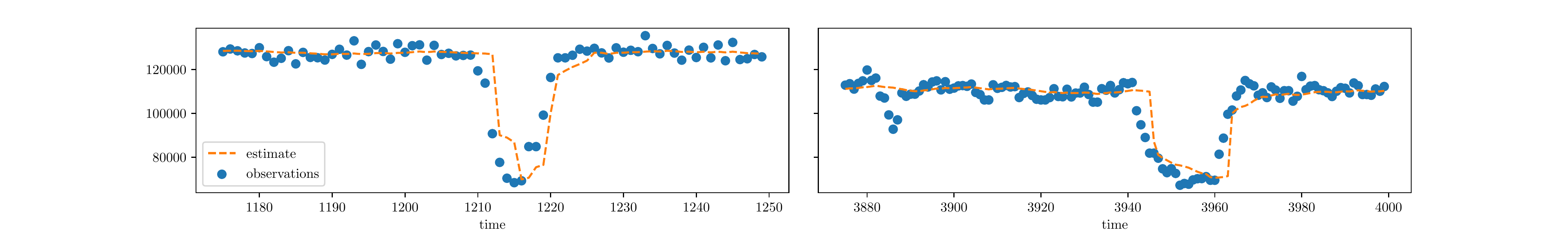}
\caption{Details of the observations and estimate on the time intervals $[1175, 1250]$ and $[3875, 4000]$.}
\label{fig:welllog_zoomin}
\end{subfigure}
\caption{Results for the change-point detection problem.}
\end{figure}

To assess the performance of the proposed solution, we consider the classical well-log data set \cite{ruanaidh2012numerical} which is a univariate time series including both change points and outliers. Within a given subset $S$ of $\Pi$, the observations are assumed to be independently sampled from $p_Y(\cdot \given \theta_S) = \text{N}(\theta_S, \sigma^2)$ with $\sigma = 2500$. The uncertain variable $\bm{\theta}$ on $\Theta = \mathbb{R}$ is therefore the mean of the observation process and we consider the uninformative prior $f_{\bm{\theta}} = \overline{\text{N}}(0, 0) = \bm{1}$, the normal possibility function $\overline{\text{N}}(\mu_t, \lambda_t)$ being conjugate for the normal likelihood. In order to take into account the deviations from the constant mean model considered for the behaviour between change points, a discount of $0.9$ is implemented for the precision $\lambda_t$, i.e.\ we consider that the precision evolves as $\lambda_t = 0.9 \lambda_{t-1} + \sigma^{-2}$; this discounting can be defined formally for possibility functions via a power. The possibility of change point $h(r)$ is assumed constant and equal to $2.5 \times 10^{-3}$, which models that there is no information about specific run-lengths and that, overall, the maximum probability of change point is smaller than $1$ in $400$ iterations. Our method considers that four of the rapid changes in the time series are change points, as opposed to the method of \cite{knoblauch2018doubly} which considers these events as outliers. Details of two of these events in Figure~\ref{fig:welllog_zoomin} show that both interpretations can be deemed correct. If the size of the jump in these events indicates that the corresponding observations are indeed outliers, then a possibility function describing the likely magnitude of the jumps can be added in the model. In terms of computational time, our method takes less than $12$ milliseconds per observation using a 2.3 GHz Intel Core i5, whereas the standard Bayesian procedure takes about $7$ milliseconds per observation; the computational complexity is however of the same order.

\section{Conclusion}
\label{sec:conclusion}

We have introduced a general method for robust Bayesian estimation in the context of possibility theory. Our method leverages the fundamental difference in the meaning of the evidence between possibility theory and probability theory. The obtained robustness complements the strengths of possibilistic inference in practical settings such as the availability of proper uninformative priors and the ability to take into account limited knowledge about some aspects of the model without introducing additional levels of hierarchy. This last point was illustrated in the change-point detection problem but applies more generally to complex models. Future work include the theoretical analysis of our method in terms of robustness and asymptotic properties.

\bibliographystyle{abbrv}
\bibliography{References}

\appendix

\section{Unbounded likelihood}
\label{sec:s_unbounded}

We first study the 1-dimensional version of the problem considered in Section~\ref{sec:extended_feature}, that is, we consider the likelihood $p_Y(\cdot \given \mu, \lambda) = \text{N}(\mu^*, 1/\lambda^*)$ where $\mu^*$ and $\lambda^*$ are respectively the mean and the precision. The challenge in this case stems from the fact that we receive observations one by one instead of receiving them in batches. The implications are that the likelihood for any given observation $y$ is unbounded when $\mu = y$ and $\lambda \to \infty$, and we have to rely on the method presented in Section~\ref{sec:unbounded_likelihood}. As in the probabilistic case, the conjugate prior for the normal likelihood with unknown mean $\bm{\mu}$ and precision $\bm{\lambda}$ is of the normal-gamma form:
\[
\overline{\text{NG}}(\mu, \lambda; \mu_0, k, \alpha, \beta) = \overline{\text{N}}(\mu; \mu_0, k\lambda) \overline{\text{G}}(\lambda; \alpha, \beta),
\]
where $\overline{\text{G}}(\alpha, \beta)$ is the gamma possibility function defined for any $(\alpha, \beta) \in \{0\} \times [0,\infty) \cup (0,\infty)^2$ as
\[
\overline{\text{G}}(\lambda; \alpha, \beta) = \bigg(\dfrac{\beta \lambda}{\alpha}\bigg)^{\alpha} \exp\big( \alpha - \beta \lambda \big).
\]
Based on the change of variable formula for possibility functions, a prior on the variance $\bm{s} = \bm{\lambda}^{-1}$ could be equivalently defined via the inverse-gamma possibility function characterised by $\overline{\text{IG}}(s; \alpha, \beta) = \overline{\text{G}}(s^{-1}; \alpha, \beta)$ with the added advantage that the expected value transforms coherently between the two, i.e.\ $\mathbb{E}^*(\bm{\lambda}) = \alpha/\beta = \mathbb{E}^*(\bm{s})^{-1}$.

We introduce a first version of the unknown parameter as $\bm{\psi} = (\bm{\mu}, \bm{\lambda})$ on $\Theta = \mathbb{R} \times (0, \infty)$ and notice that $L( \psi \given y)$ is well defined when $f_{\bm{\psi}} = \overline{\text{NG}}(\mu_0, 0, 0, \beta)$ with $\beta > 0$, $\mu_0$ being irrelevant. Indeed, having $\beta > 0$ leads to an exponential decay in $\lambda$ which prevents the likelihood from growing without bounds. Now introducing the second version $\bm{\theta}$ of the unknown parameter as in the main article, we define the prior on $\bm{\theta}$ as $f_{\bm{\theta}} = \overline{\text{NG}}(\mu_0, 0, 0, \beta_0)$ with $\beta_0 > 0$.

The prior being defined, we can describe the general recursion at iteration $t \in \{1,\dots,T\}$. Instead of considering a split of the overall posterior information about $\bm{\theta}$ into two parts, as suggested for the general case in Section~\ref{sec:unbounded_likelihood}, we consider a more careful approach and only split the information about the precision by defining
\begin{align*}
f^{(\omega)}_{\bm{\theta}}(\mu, \lambda \given y_{1:t-1}) & \doteq \overline{\text{N}}(\mu; \mu_{t-1}, k_{t-1} \lambda) \overline{\text{G}}(\lambda; \omega\alpha_{t-1}, \omega\beta_{t-1}) \\
f^{(\omega)}_{\bm{\psi}}(\mu, \lambda \given y_{1:t-1}) & \doteq \overline{\text{G}}(\lambda; (1-\omega)\alpha_{t-1}, (1 - \omega) \beta_{t-1})
\end{align*}
for some $\omega \in (0,1)$, where $\mu_{t-1}$, $k_{t-1}$, $\alpha_{t-1}$ and $\beta_{t-1}$ are the updated parameters at iteration $t-1$. We then proceed with the recursion as
\begin{equation*}
f^{\mathrm{r}}_{\bm{\theta}}(\theta \given y_{1:t}) \doteq \dfrac{L_t(\theta \given y_t)^{\gamma_t} f^{(\omega)}_{\bm{\theta}}(\theta \given y_{1:t-1})}{\sup_{\theta' \in \Theta} L_t(\theta' \given y_t)^{\gamma_t} f^{(\omega)}_{\bm{\theta}}(\theta' \given y_{1:t-1})}
\end{equation*}
where $\gamma_t = \sup_{\theta \in \Theta} p_Y(y_{t} \given \theta) f^{(\omega)}_{\bm{\psi}}(\theta \given y_{1:t-1})$ and
\begin{equation*}
L_t(\theta \given y_t) = \gamma_t^{-1} p_Y(y_{t} \given \theta) f^{(\omega)}_{\bm{\psi}}(\theta \given y_{1:t-1}),
\end{equation*}
where the dependence of $L_t(\cdot \given y_t)$ on $y_{1:t-1}$ is omitted for the sake of simplicity. We emphasise once more that, although both the prior and the likelihood depend on the previous observations, there is no double counting of information thanks to the two-way splitting of the posterior at the previous iteration. The considered model leads to the consistency
\[
c(y_t) = \bigg( \dfrac{(1-\omega)\beta_{t-1}}{(1-\omega)\alpha_{t-1}+\nicefrac{1}{2}} \bigg)^{(1-\omega)\alpha_{t-1}+\nicefrac{1}{2}} \bigg( \dfrac{\beta_{t-1}}{\alpha_{t-1}} \bigg)^{\omega\alpha_{t-1}} \bigg( \dfrac{\alpha_{t-1} + \nicefrac{1}{2}}{\hat{\beta}_t} \bigg)^{\alpha_{t-1} + \nicefrac{1}{2}},
\]
with
\[
\hat{\beta}_t = \beta_{t-1} + \dfrac{k_{t-1}}{2(k_{t-1}+1)} (\mu_{t-1} - y_t)^2.
\]
Since there are two parameters in the model, we define the discount as $\gamma_t = \sqrt{c(y_t)}$. In particular, one can check that if $\omega = k_{t-1} = 0$ then $\gamma_t = 1$; indeed, these prior parameters imply that the prior $f^{\sb{\omega}}_{\bm{\theta}}(\cdot \given y_{1:t-1})$ is uninformative. In practice, $\omega$ should not be set to $0$ since this would yield the standard (non-robust) posterior. The update for the parameters of the model is as follows:
\begin{align*}
\mu_t & = \dfrac{k_{t-1} \mu_{t-1} + \gamma_t y_t}{k_{t-1} + \gamma_t} \\
k_t & = k_{t-1} + \gamma_t \\
\alpha_t & = \alpha_{t-1} + \gamma_t \Big( (1-\omega)\alpha_{t-1} + \dfrac{1}{2} \Big) \\
\beta_t & = \beta_{t-1} + \dfrac{\gamma_t k_{t-1}}{2(k_{t-1} + \gamma_t)} (\mu_{t-1} - y_t)^2.
\end{align*}
The derivation of these formulas follows from the usual calculations and from $\overline{\text{G}}(\alpha,\beta)^{\omega} = \overline{\text{G}}(\omega\alpha,\omega\beta)$. 

\begin{figure}
\centering
\includegraphics[trim=95pt 5pt 115pt 35pt,clip,width=\textwidth]{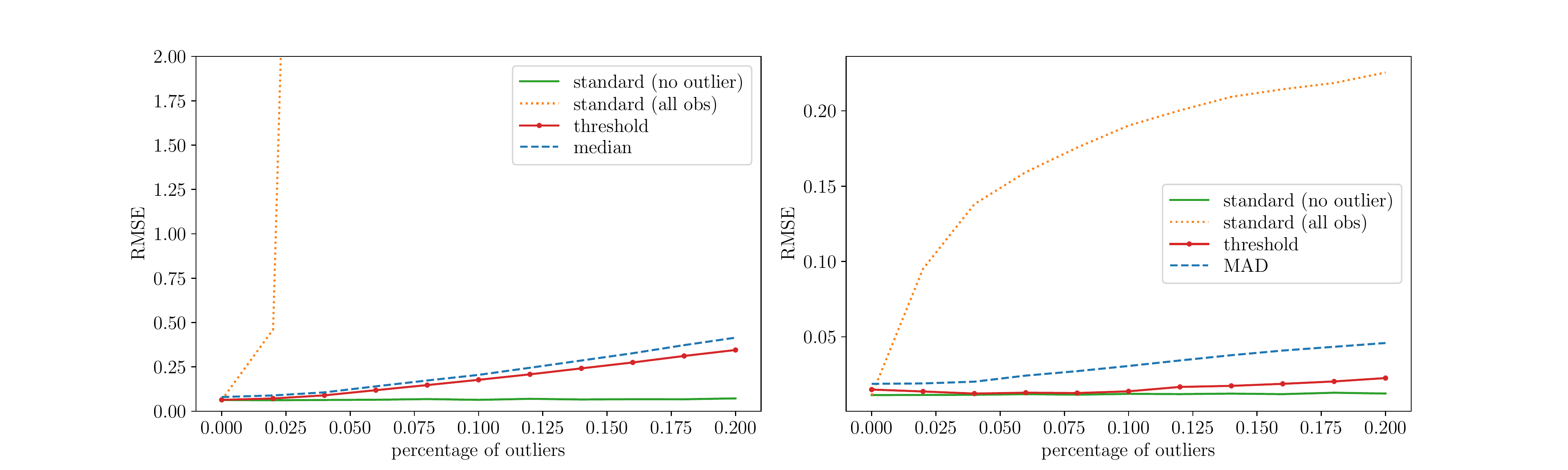}
\caption{RMSE in the mean $\mu$ (left) and in the precision $\lambda$ (right) as a function of the probability of outliers ($\epsilon$) for different methods, averaged over $1000$ repeats, for the model of Section~\ref{sec:s_unbounded}.}
\label{fig:normal_gamma}
\end{figure}

To illustrate the behaviour of the proposed approach on this model, we consider the case where $\mu^* = 2$, $\lambda^* = \nicefrac{1}{4}$, $\beta_0 = 1$ and $\omega = 0.99$. The probability of outlier is denoted $\epsilon$ and each outlier is sampled from a standard Cauchy distribution. The alternative approach of Section~\ref{sec:alternative} is considered with a threshold $\tau$ equal to $0.1$. We focus on this approach since discounting each observation would lead to a positive bias in the precision due to inliers in the tail of the distribution being more strongly discounted than other inliers. This effect arises because of the limited information about the precision in a single observation, similarly to the case of Example~\ref{ex:Bernoulli}. Figure~\ref{fig:normal_gamma} shows the evolution of the root mean square error (RMSE) in the estimation of the two parameters of interest as a function of $\epsilon$. In addition to the standard results obtained by applying Bayesian inference to all the observations or to the inliers only, we display one of the standard ways of obtaining robust estimate of location and scale, i.e., the median and the median absolute deviation (MAD) respectively. In spite of being averaged over $1000$ repeats, the error in location for Bayesian inference based all the observations displays a very erratic behaviour due to the Cauchy distribution being heavy-tailed; the corresponding graph is partially cropped in the left panel of Figure~\ref{fig:normal_gamma} in order to highlight the difference between the other methods. The performance of the proposed approach is particularly good for the precision parameter.

\begin{figure}
\centering
\includegraphics[trim=20pt 5pt 55pt 35pt,clip,width=0.5\textwidth]{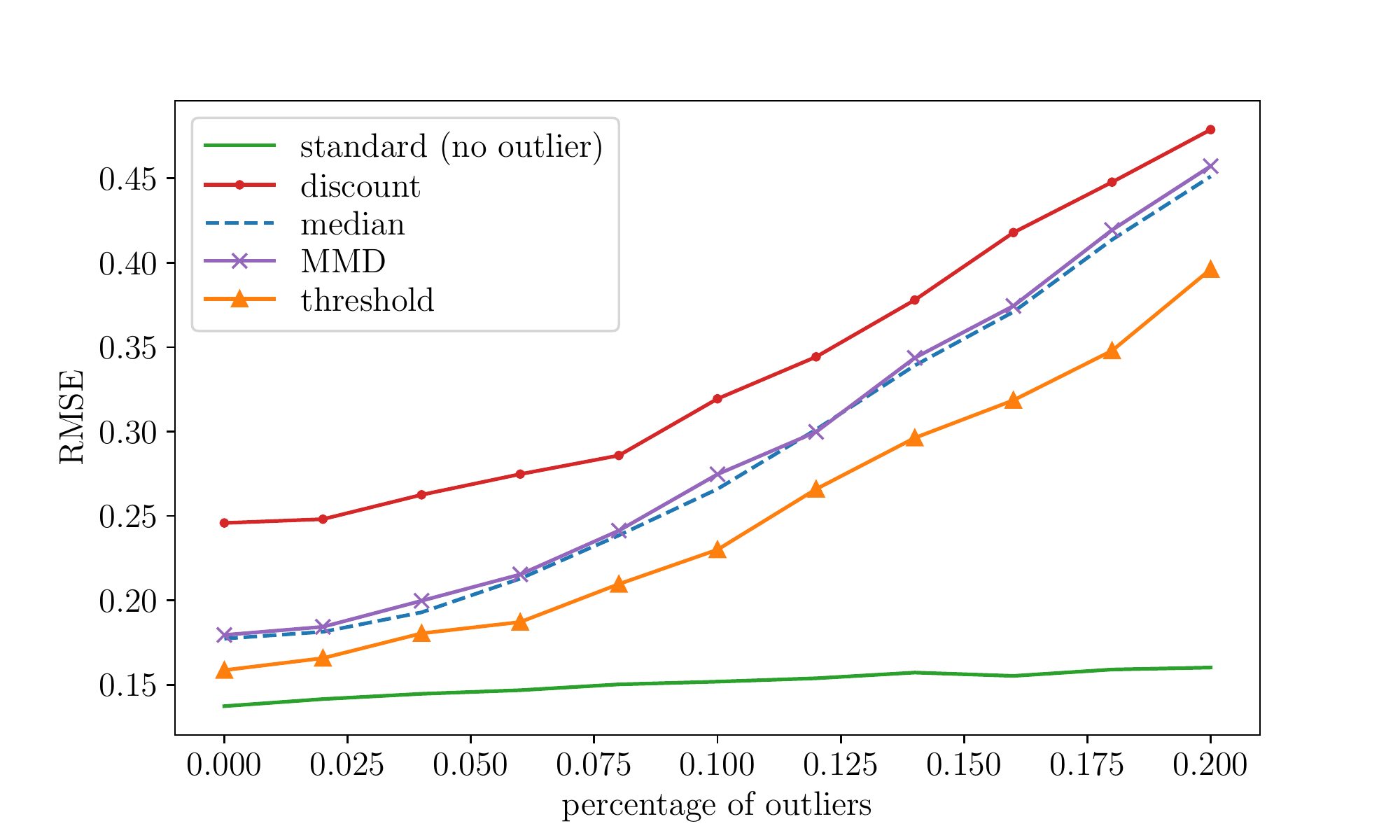}
\caption{RMSE in the mean $\mu$ as a function of the probability of outliers ($\epsilon$) for different methods, averaged over $1000$ repeats, for the normal model of Section~\ref{sec:s_unbounded}.}
\label{fig:normal}
\end{figure}

In order to compare with methods based on maximum mean discrepancy (MMD) \cite{cheriefabdellatif2020mmd}\footnote{We would like to thank the authors of \cite{cheriefabdellatif2020mmd} for kindly sharing their code with us.}, we also consider the simplified model where the precision of the observations is known. Under the same conditions as above, we compute the RMSE for all methods and display it in Figure~\ref{fig:normal} as a function of the probability of outliers. While the ``discount'' approach can be applied in this case, its performance is lesser than the other robust methods. The alternative based on a threshold $\tau = 0.1$ allows for improving the performance. While, in this particular scenario, the MMD-based approach displays a similar performance to the median, it can be applied in more general setting including uniform likelihoods as in the following section. The average run-times per repeat on an 1.8 GHz Intel Core i7 were of the order of $1$ milliseconds for the median and standard Bayesian approach, of the order of $10$ milliseconds for the proposed approach and of the order of $10$ seconds for the MMD-based approach.

\section{Likelihood with observation-dependent support}
\label{sec:s_observation-dependent}

As discussed in Section~\ref{sec:observation-dependent_support}, the proposed approach cannot take into account outliers that yield a credibility of $0$ in the region of the true parameter, since a positive discount will have no effect on such a likelihood. In order to illustrate this aspect, we consider the uniform likelihood $p_Y(\cdot \given \theta) = \theta^{-1}\bm{1}_{[\cdot, \infty)}(\theta)$ and multiple i.i.d.\ observations $y_1,\dots,y_T$. We denote the unknown parameter by $\bm{\theta}$ and assume it to be described by the Pareto possibility function on $\Theta = (0,\infty)$ defined as
\[
\overline{\text{Pa}}(\theta; \alpha, s) = \bigg( \dfrac{s}{\theta} \bigg)^{\alpha} \bm{1}_{[s, \infty)}(\theta),
\]
for some $s, \alpha \geq 0$, which verifies $\mathbb{E}^*(\bm{\theta}) = s$. The Pareto possibility function is the conjugate prior for the uniform likelihood; we note that the parameters of $f_{\bm{\theta}}(\cdot \given y_{1:t-1}) = \overline{\text{Pa}}(\alpha_{t-1}, s_{t-1})$ are $s_{t-1} = \max_{t' \in \{1,\dots,t-1\}} y_{t'}$ and $\alpha_{t-1} = t-1$, and we compute the consistency of $y_t$ as
\[
c(y_t) =
\begin{cases}
y_{t-1} / s_{t-1} & \text{if } y_{t-1} < s_{t-1} \\
s_{t-1}^{\alpha_{t-1}} / y_t^{\alpha_{t-1}} & \text{otherwise}.
\end{cases}
\]
The expression of $c(y_t)$ when $y_{t-1} \geq s_{t-1}$ is the most useful for our purpose since it evaluates how likely is the jump from $s_{t-1}$ to $y_{t-1}$ given the number of observations already received. We therefore assume that the observations are ordered, i.e.\ $y_1 \leq \dots \leq y_T$. However, raising $L(\theta \given y_t) = \overline{\text{Pa}}(\theta; 1, y_t)$ to the power $c(y_t)$ still leads to a posterior that is supported by $[y_t,\infty)$, which is not desirable if $y_t$ is an outlier; this is because the indicator function in $L(\theta \given y_t)$ cannot be discounted.

\begin{figure}
\centering
\begin{subfigure}[b]{0.48\textwidth}
\includegraphics[trim=30pt 5pt 55pt 35pt,clip,width=\textwidth]{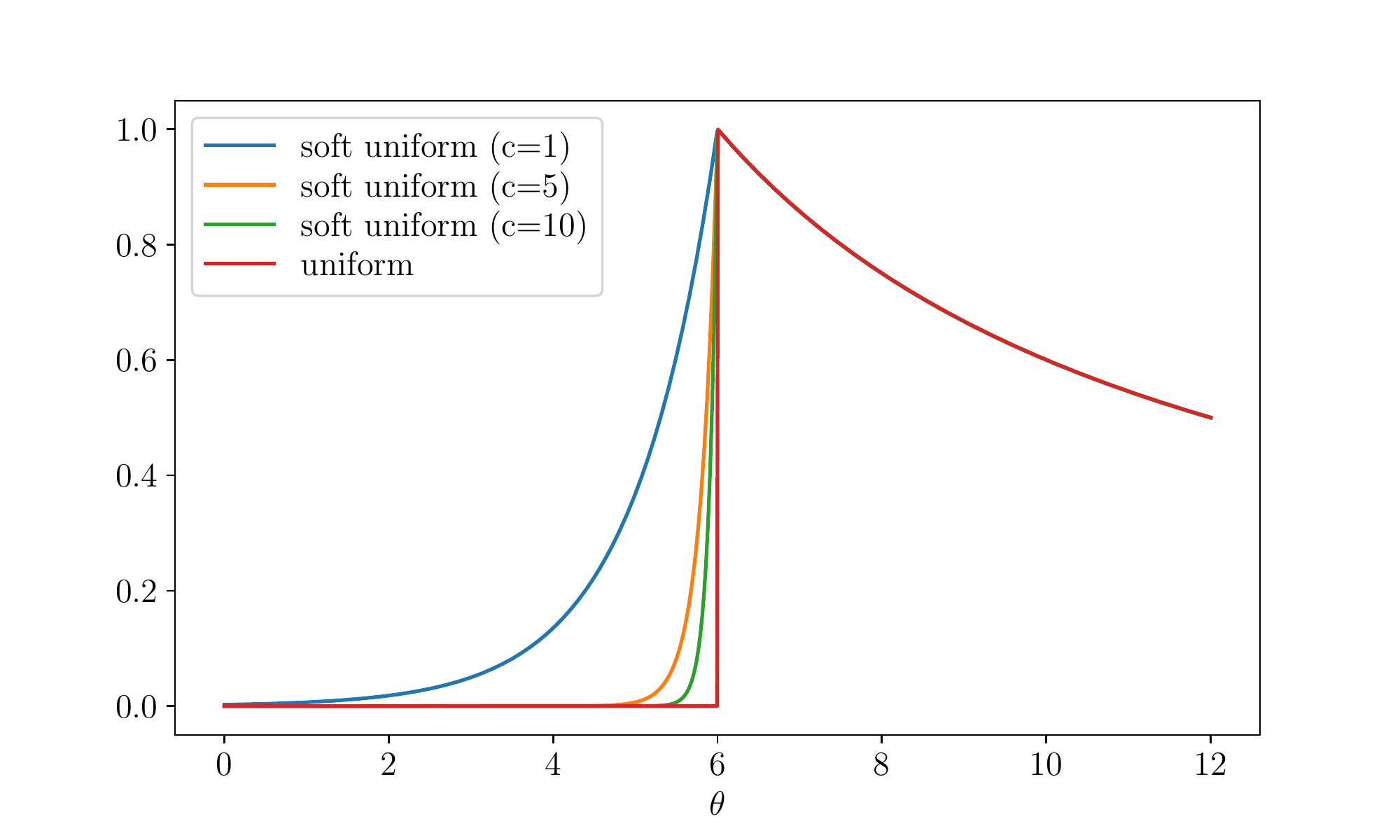}
\caption{Soft-uniform likelihood for different values of the exponential decay $c$.}
\label{fig:soft-uniform}
\end{subfigure}\quad
\begin{subfigure}[b]{0.48\textwidth}
\includegraphics[trim=30pt 5pt 55pt 35pt,clip,width=\textwidth]{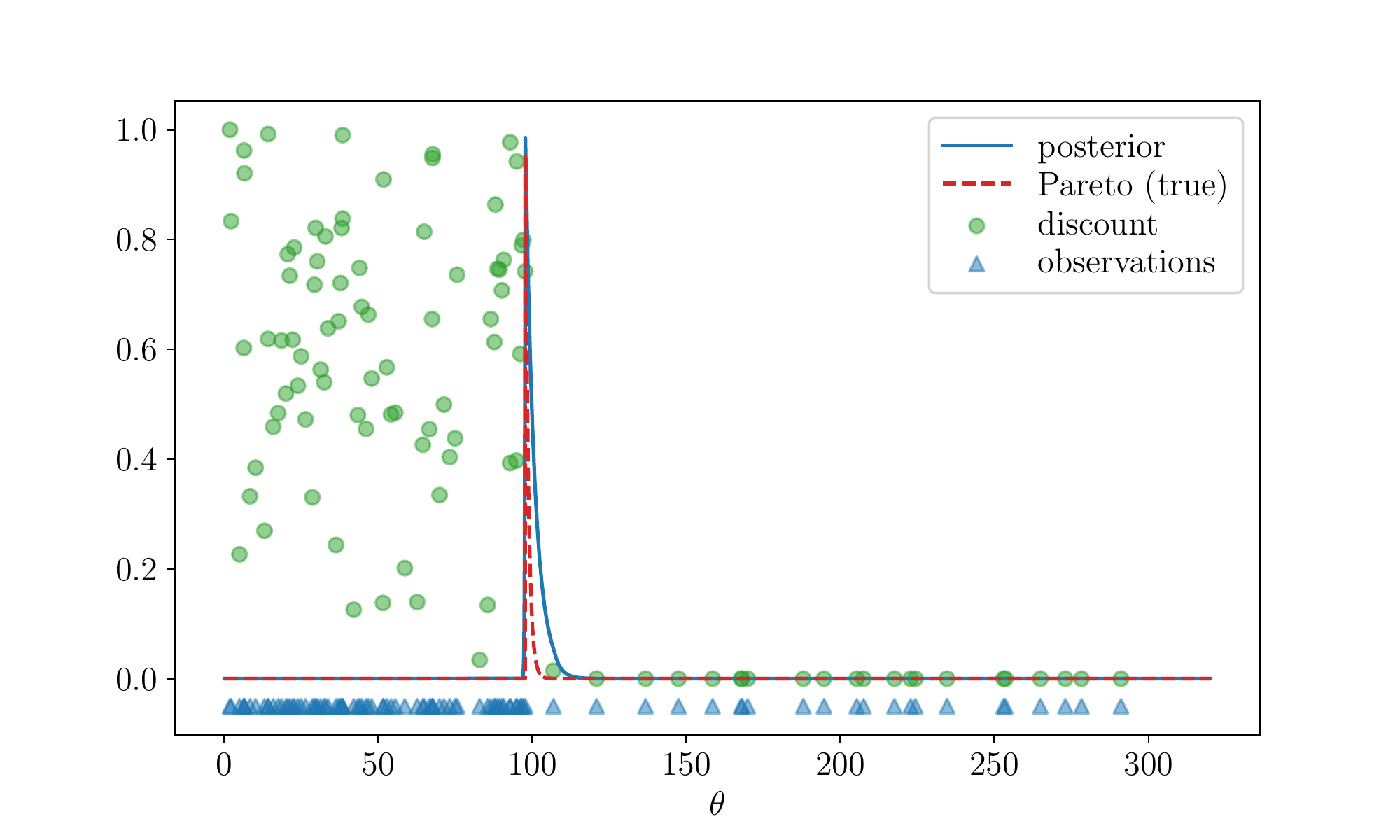}
\caption{Results with soft-uniform likelihood where both robust and standard posteriors.}
\label{fig:soft-uniform_results}
\end{subfigure}
\caption{Likelihood and posterior for the model of Section~\ref{sec:s_observation-dependent}.}
\end{figure}

To allow for the proposed approach to be applied, we introduce a ``soft'' version of $L(\theta \given y_t)$ as
\[
\tilde{L}(\theta \given y_t) = \dfrac{y_t}{\theta \lor y_t} \exp\big( c (\theta - \theta \lor y_t) \big)
\]
with $c$ the coefficient of the exponential decay and $\lor$ the maximum as a binary operator (assumed to have lower precedence than multiplication but higher precedence than addition). The possibility function $\tilde{L}(\cdot \given y_t)$ is plotted in Figure~\ref{fig:soft-uniform} for different values of $c$.

Denoting $\gamma_{t'}$ the discount at iteration $t' \in \{1,\dots,T\}$, the unnormalised robust posterior possibility functions (with discounted likelihood) at iteration $t-1$ is of the form
\[
\tilde{f}_{\bm{\theta}}(\theta | y_{1:t-1}) = \exp\bigg( c \sum_{t'=1}^{t-1} \gamma_{t'} (\theta - \theta \lor y_{t'}) \bigg) \prod_{t'=1}^{t-1} (\theta \lor y_{t'})^{-\gamma_{t'}},
\]
We then define $f^{\mathrm{r}}_{\bm{\theta}}(\cdot | y_{1:t-1}) = C(y_{1:t-1}, \gamma_{1:t-1})^{-1} \tilde{f}_{\bm{\theta}}(\cdot | y_{1:t-1})$ as the normalised robust posterior possibility function. The normalising constant is not known analytically, but the maximum defining it must be achieved at one of the observations and is therefore easily computable as
\[
C(y_{1:t-1}, \gamma_{1:t-1}) = \max_{t' \in \{1,\dots,t-1\}} \tilde{f}_{\bm{\theta}}(y_{t'} | y_{1:t-1}).
\]
The discount $\gamma_t$ of the observation $y_t$ is then defined as
\[
\gamma_t = c(y_t) = \sup_{\theta \in \Theta} \tilde{L}(\theta \given y_t) f_{\bm{\theta}}(\theta | y_{1:t-1}) = y_t \dfrac{C(y_{1:t}, (\gamma_{1:t-1},1))}{C(y_{1:t-1}, \gamma_{1:t-1})}.
\] 
The considered prior/posterior possibility functions cannot be considered as conjugate since their complexity increases with the number of observations. Yet, they can be computed exactly with a complexity of the order of $t$, which is easily achievable even for large values of $t$. The main limitation of this model is that it introduces an additional tuning parameter $c$, which must be suitably chosen. Finding automatic ways to set $c$ depending on prior information on the problem at hand will be the topic of future work.

The proposed solution is illustrated in a scenario where the true parameter is $\theta^* = 100$ and where there are $T = 100$ observations. The probability of outlier is $\epsilon = 0.25$ and outliers are sampled from a normal distribution with mean $200$ and standard deviation $50$. The results for one specific realisation of the observations are displayed in Figure~\ref{fig:soft-uniform_results}, where both the robust posterior and the standard posterior based on inliers only are included. This is a challenging case where the outliers are close to the inliers. The coefficient $c$ of the exponential decay is set to $10$ in this scenario; however this parameter depends on both $\theta^*$ and $T$ in ways that remain to be explored. The difficulty of the scenario means that, in some cases, the estimate $\mathbb{E}^*(\bm{\theta} \given y_{1:T})$ will be either much smaller or much larger than the true parameter $\theta^*$. This can be seen by comparing the RMSE and the median absolute error over $1000$ repeats, which were found to be equal to $17.7$ and $1.27$ respectively.

\end{document}